\documentclass[twocolumn]{revtex4-1}

\usepackage{graphicx}
\usepackage{color}
\begin{document}

\title{Pinning of Ferroelectric Bloch Wall at a Paraelectric Layer}

\author{V. Stepkova}
\author{J. Hlinka}
\affiliation{Institute of Physics, The Czech Academy of Sciences, Prague, Czech Republic}

\date{\today}

\begin{abstract}
Phase-field simulations of domain walls between antiparallel ferroelectric domains in BaTiO$_3$-SrTiO$_3$ crystalline superlattices suggest that a paraelectric layer with a thickness comparable to the thickness  of the domain wall itself can act as an efficient pinning layer. Moreover, such layer is capable to unwind the characteristic structure of a ferroelectric Bloch wall passing through it.
\end{abstract}


\maketitle 

\section{Introduction}

Nanometer scale mixtures of paraelectric and ferroelectric materials in disordered solid solutions or in very fine artificial crystalline superlattices are often showing qualitatively similar domain phenomena as the parent ferroelectric materials. For example, in case of BaTiO$_3$-SrTiO$_3$ superlattices with few atomic layers of SrTiO$_3$ only, domain walls are simply expected to penetrate through BaTiO$_3$/SrTiO$_3$ interfaces.  In general, it can be expected that small amount of paraelectric defects, smaller or thinner than the correlation length, will not substantially alter the superposed domain structure in a strong ferroelectric material like BaTiO$_3$. In simple words, a sufficiently thin paralectric layer is effectively polarized by the neighboring ferroelectric material.

However, little is known about how robust is the inner polarization, present within the nanoscale thickness of ferroelectric Bloch walls \cite{Marton2010, Taherinejad2012}. For example, it was not yet clarified whether such a localized polarization sustain the chemical stoichiometry concentration fluctuations typical for example for relaxor ferroelectric perovskites. Similarly, we are not aware of any device geometries allowing to define or alter the helicity of Bloch walls.

It seems that the basic model situation is that of the interaction of ferroelectric Bloch wall with a layer of paraelectric material. It is intuitively clear that paraelectric layer would act as pinning loci for the Ising ferroelectric wall,
because the inside of the Ising wall is not polarized anyway. In that respect, the largest effect is expected for a layer with thickness matching that of the domain wall. 

In the model case studied here, a nanometer thick SrTiO$_3$ layer intercalated in
BaTiO$_3$ was considered. It is found that such a layer acts as a pinning loci not only for the Ising wall, but also for the Bloch wall. Moreover, our results suggest that such pinned Bloch wall can loose practically all its inner polarization. This result could be perhaps used to set or modify helicity of Bloch walls passing through conveniently placed paraelectric gate layers in future domain wall-based devices.

\section{Technical details}

To estimate the interaction of ferroelectric Bloch walls with a nanoscale paralectric layer of similar material, we have considered  a hypothetical BaTiO$_3$-SrTiO$_3$ crystalline superlattice, formed by 1\,nm thin SrTiO$_3$ paraelectric layers separated by about 13\,nm thick BaTiO$_3$ ferroelectric slabs. The SrTiO$_3$ layers were normal to [$\bar{2}$11] crystallographic direction, common to the  parent cubic lattice of both BaTiO$_3$ and SrTiO$_3$. Whole structure fit in a 64$\times$128$\times$128 equidistant point simulation box with its  the edges along the principal pseudocubic crystallograpic axes and with 0.5\,nm lattice step.
According to the ab-initio calculations \cite{Taherinejad2012} and Ginzburg-Landau-Devonshire model, the 180-degree ferroelectric walls of this orientation in  rhombohedral BaTiO$_3$ are Bloch like. 

The BaTiO$_3$ and SrTiO$_3$ Ginzburg-Landau-Devonshire model potential parameters used in the present calculations are those of Ref.\,\cite{StepkovaPRB14}, except for the temperature parameter, with was set to 118\,K here (low temperature is needed  to drive BaTiO$_3$ into the rhombobedral ferroelectric phase).  Phase-field simulations for pure BaTiO$_3$ single crystal
and for the BaTiO$_3$-SrTiO$_3$ crystalline superlattice at stress-free mechanical conditions were performed using the phase field simulation code ferrodo \cite{Marton2006}. The approximate profile of the ferroelectric Bloch wall is known from the previous calculation and this knowledge could be conveniently used for setting the initial conditions for the present phase field simulations.
In particular, the initial conditions were set in a way to favor the orientation, distance and the location of the walls in the simulation box, but there were no real constrains introduced there, only the natural limitations resulting from the standard protocol of simulated annealing procedure, governed by the time-dependent Ginzburg-Landau equation.

\section{Results}

\begin{figure}[ht]
\includegraphics[width=8cm]{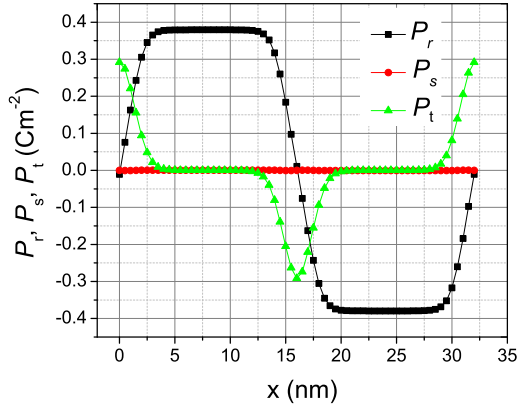}
\caption{ Relaxed profile of the polarization components across 180-degree ferroelectric  wall in rhombohedral BaTiO$_3$ .}
\label{NewFigBlochBaTiO$_3$}
\end{figure}

The peculiarity of the investigated ferroelectric domain wall is best understood when the polarization is expressed in the symmetry adapted Cartesian system \cite{Marton2010} associated with the set of orthogonal unit
vectors ${\bf r} \parallel [111]$, ${\bf s} \parallel [\bar{2}11]$, and  ${\bf t} \parallel [0\bar{1}1]$. By definition \cite{Marton2010}, the adjacent domains differ in the sign of the $P_{\rm r}$ component.  The relaxed, equilibrium polarization profile in the simulation for pure BaTiO$_3$ is shown in  Fig.\,\ref{NewFigBlochBaTiO$_3$}.

\begin{figure}[ht]
\includegraphics[width=8cm]{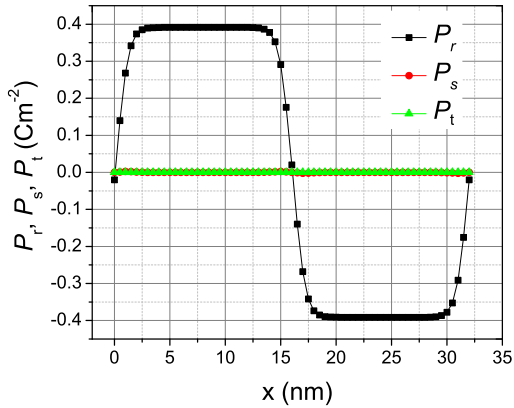}
\caption{ Relaxed profile of the polarization components across 180-degree ferroelectric  wall in rhombohedral BaTiO$_3$ under epitaxial compressive stress of 3GPa. }
\label{labelfig2}
\end{figure}

In case of Ising walls, the integral of the $P_{\rm t}$ component across a given wall is zero.
In case of Bloch walls,  there is also an overall polarization in the $P_{\rm t}$ component  within the few nm thickness of the give domain wall itself. This inner polarization can be negative or positive. Such Bloch walls are indeed found in Fig.\,\ref{NewFigBlochBaTiO$_3$}. Alternating signs of $P_{\rm t}$ component in subsequent domain walls indicate that the energetically equivalent Bloch walls present in these  calculations have the same helicity. For comparison, Fig.\,\ref{labelfig2} shows the profile of the Ising domain wall, which is obtained by the same simulation but under an epitaxial compressive stress of 3.0\,GPa, applied in the plane perpendicular to the spontaneous polarization as described in  Ref.\,\cite{StepkovaJPCM}.

\begin{figure}[ht]
\includegraphics[width=8cm]{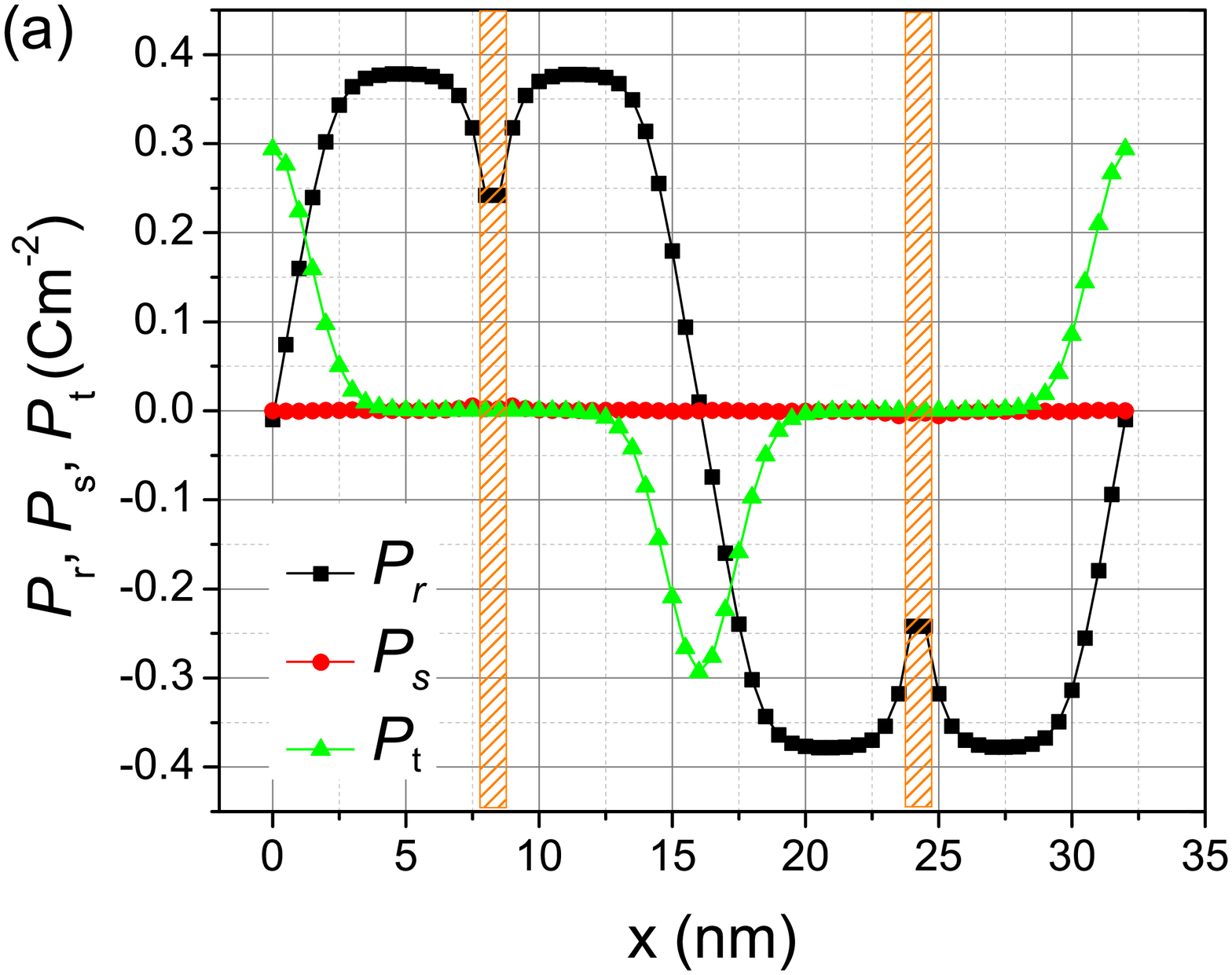}
\includegraphics[width=8cm]{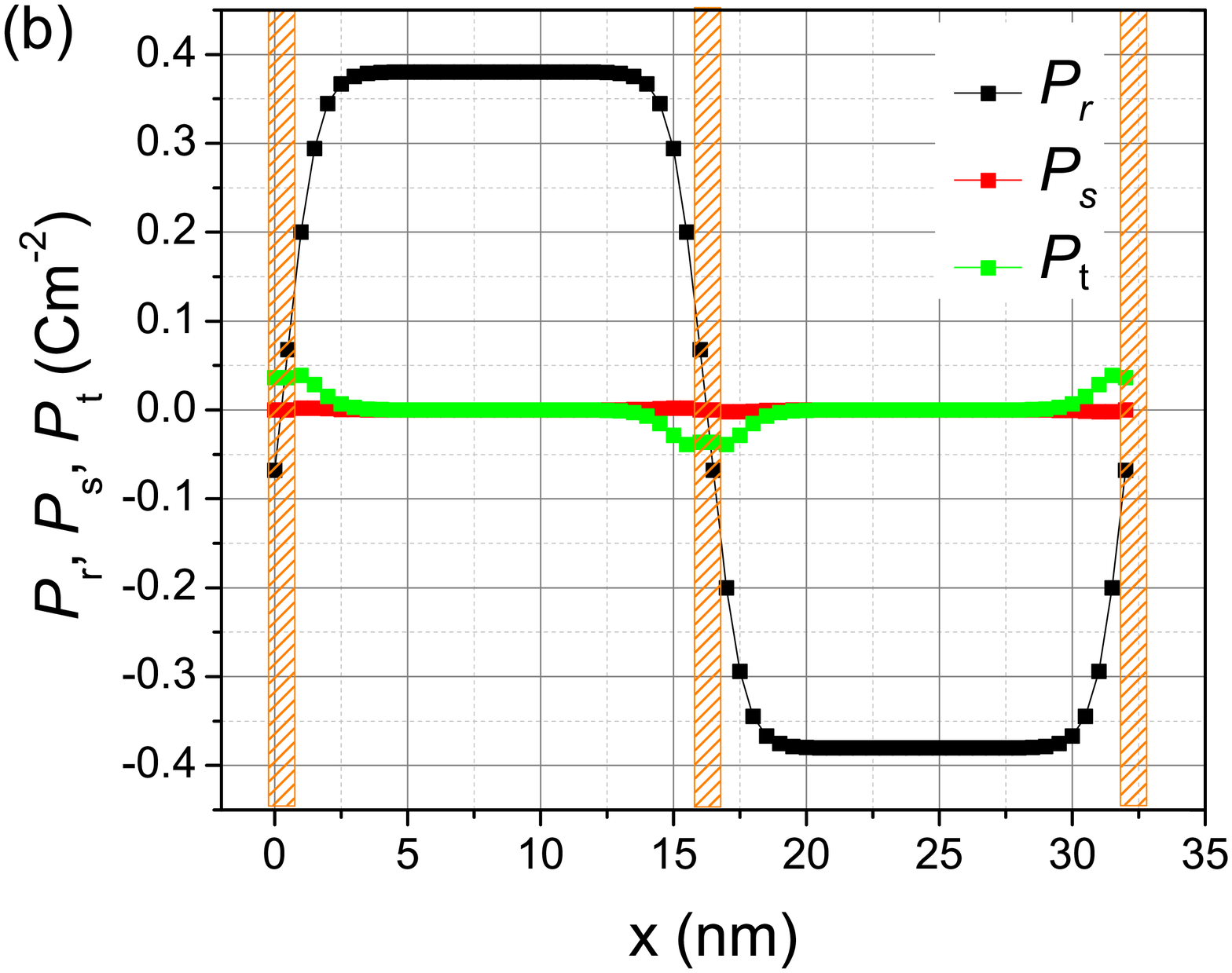}
\caption{ Relaxed profile of the polarization components across 180-degree ferroelectric  walls in rhombohedral BaTiO$_3$ intercalated with 1\,nm thick layers of 
SrTiO$_3$. (a) SrTiO$_3$ layer within the ferroelectric domain (b) SrTiO$_3$ layer located at the ferroelectric domain wall.} 
\label{NewFigSUPERLATTICE}
\end{figure}

The relaxed polarization profile in the superlattice containing SrTiO$_3$ layers is shown in Fig.\,\ref{NewFigSUPERLATTICE}. When the domain wall is far away from the SrTiO$_3$ layer, the domain wall profile is barely modified, only the spontaneous polarization in the domain is somewhat reduced (by about 40 percent, see Fig.\,\ref{NewFigSUPERLATTICE}a). In contrast, when the domain wall happens to be located right at the SrTiO$_3$ layer, the $P_{\rm t}$ component is suppressed considerably (by more than 80 percent in Fig.\,\ref{NewFigSUPERLATTICE}b). In other words, its Bloch character is strongly suppressed. Moreover, 
when the domain wall is initially located at few nm away the SrTiO$_3$ layer, it moves towards the SrTiO$_3$ layer, so that the position of the SrTiO$_3$ layer is acting as a pinning center.

\begin{figure}[ht]
\includegraphics[width=8cm]{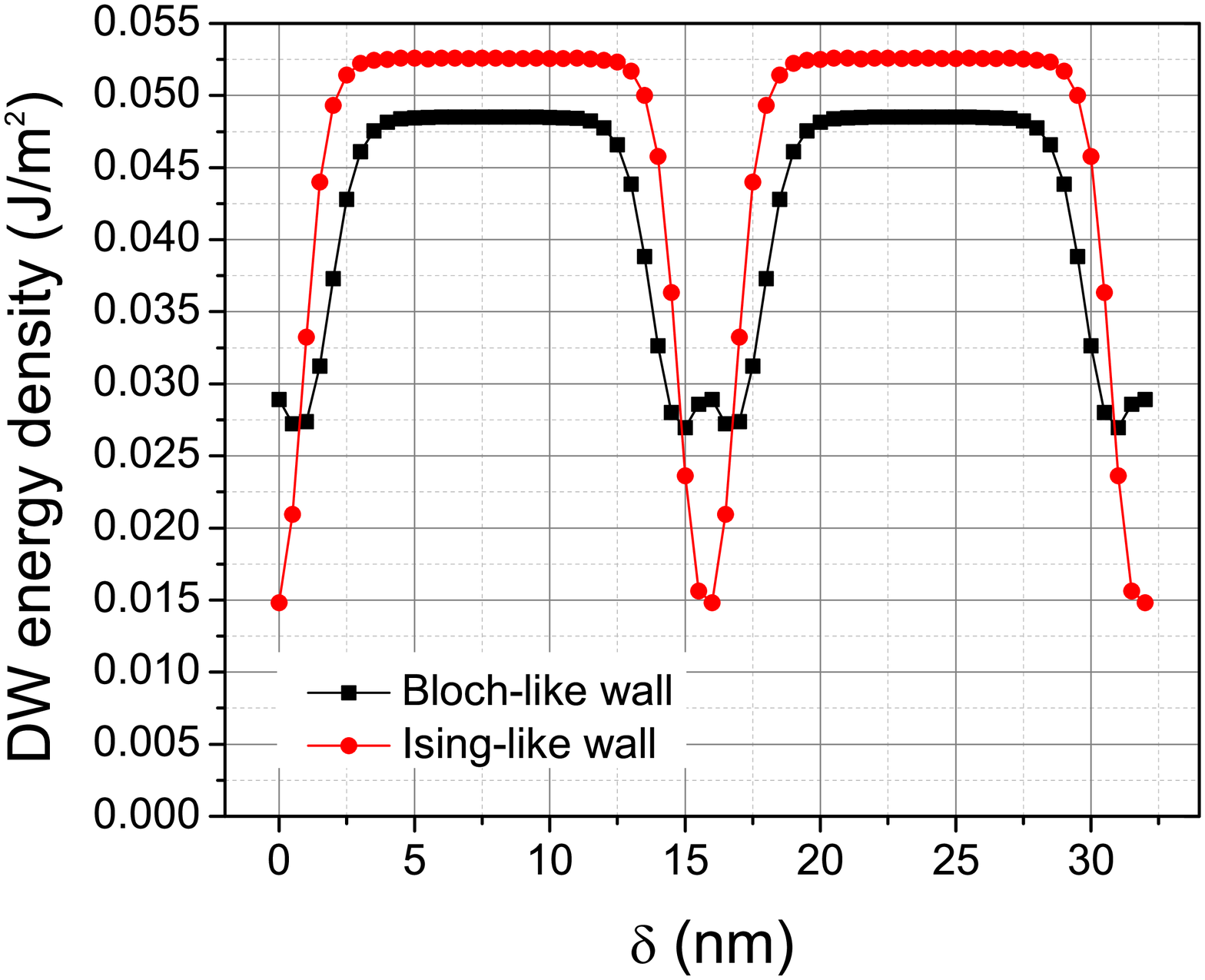}
\caption{ Landau part of the domain wall energy density as a function of the position of the ferroelectric domain wall, estimated by sliding a rigid polarization profile across the potential relief of the superlattice described in the text.}
\label{NewFig3}
\end{figure}

In order to estimate qualitatively the hight and shape of the potential well formed by the SrTiO$_3$ layer, we have also calculated the Landau energy contribution the domain wall energy density as a function of the position of the domain wall, assuming that the shape of the profile of the domain wall is not modified  while sliding across the SrTiO$_3$ layer. Result of this calculation, obtained numerically by integrating Landau energy density is shown in  Fig.\,\ref{NewFig3}. The calculation was made for the ideal profile of the Ising wall and Bloch wall of Fig.\,\ref{labelfig2} and Fig.\,\ref{NewFigBlochBaTiO$_3$}. The depth of the energy well is more shallow in case of Bloch wall but still comparable (about twice in this rough estimation).

\section{Discussion and Conclusion}

These simulations suggest that about 1\,nm thin SrTiO$_3$ layer, incorporated in ferroelectric BaTiO$_3$ crystal, is capable to influence considerably
structure and properties of the 180-degree Bloch domain walls parallel to it. Since the domain wall has lower energy when located right at the SrTiO$_3$ layer, one can speculate that a pair of such layers  can be used as
a nucleation center for favoring antiparallel ferroelectric domain with desirable crystallographic orientation of adjacent domain walls. For example, one can choose the [$\bar{2}$11] crystallographic direction, favorable for Bloch wall formation. It can also be used to pin domain walls already present in the material. Most interestingly, we have seen that the inner polarization of the Bloch wall  would be substantially reduced while passing through the thin paraelectric layer. Thus, the layer acts as a bottleneck for the helicity order parameter of the wall.

Moreover, since the Bloch character is strongly suppressed when the domain wall is right at the SrTiO$_3$ layer, the layer  can  facilitate selection of the sign of the $P_{\rm t}$ component and, therefore, selection of the sign of its helicity. In fact, the thickness of the SrTiO$_3$ layer can  be tuned in a way that the wall passing there is effectively in the state just below the phase transition from the Bloch to the Ising state. Then, as the domain wall would go through such paraelectric layer, it should easily acquire the $P_{\rm t}$ component favored
 by even quite a moderate $E_{\rm t}$ electric bias. We believe that these findings can be inspiring for design of functional properties of ferroelectric nanostructures.

\begin{acknowledgments}
This work was supported by the Czech Science Foundation (project no. 15-04121S). Authors warmly acknowledge long-term maintenance and user support of ferrodo code by Pavel Marton.
\end{acknowledgments}

\end{document}